\documentclass[twocolumn,prb,aps,showpacs,10pt]{revtex4-1}
\usepackage{amsfonts}
\usepackage{amsmath}
\usepackage{amssymb}
\usepackage{dsfont}
\usepackage{graphicx}
\usepackage{bm}
\usepackage[usenames,dvipsnames,svgnames,table]{xcolor}
\usepackage{tabularx}
\newcommand{\mbf}[1]{\mathbf{#1}}
\newcommand{\ddt}[0]{\frac{\partial}{\partial t}}

\renewcommand{\t}[1]{\textrm{#1}}
\renewcommand{\k}[0]{\mbf{k}}

\newcommand{\nn}[0]{\nonumber\\}

\newcommand{\K}[6]{K_{l#1n#3\k#5}^{l#2n#4\k#6}}
\newcommand{\Q}[6]{Q_{l#1n#3\k#5}^{l#2n#4\k#6}}

\newcommand{\C}[3]{C_{l#1\k#3}^{l#2}}

\newcommand{\Cq}[4]{\bar{C}_{l#1\k#3}^{l#2\k#4}}

\newcommand{\M}[2]{M_{n#1}^{n#2}}

\newcommand{\Cstd}[0]{\C{_1}{_2}{_1}}

\newcommand{\Cqstd}[0]{\Cq{_1}{_2}{_1}{_2}}
\newcommand{\Mstd}[0]{\M{_1}{_2}}
\newcommand{\Kstd}[0]{\K{_1}{_2}{_1}{_2}{_1}{_2}}
\newcommand{\Qstd}[0]{\Q{_1}{_2}{_1}{_2}{_1}{_2}}
\newcommand{\bstd}[0]{{b_{l_1n_1\k_1}^{l_2n_2\k_2}}}

\newcommand{\Jsd}[0]{J_{sd}}

\newcommand{\nMn}[0]{n_{Mn}}
\newcommand{\NMn}[0]{N_{Mn}}
\newcommand{\spar}[0]{{s^\|_{\k_1}}}
\newcommand{\sper}[0]{{s^\perp_{\k_1}}}
\begin{document}
\title{Comparison between a quantum kinetic theory of spin transfer dynamics in Mn doped bulk semiconductors and its Markov limit for non-zero Mn magnetization}
\author{M.~Cygorek}
\author{V.~M.~Axt}
\affiliation{Theoretische Physik III, Universit{\"a}t Bayreuth, 95440 Bayreuth, Germany}
\begin{abstract}
We investigate the transfer between carrier and Mn spins due to the s-d-exchange interaction 
in a Mn doped bulk semiconductor within a microscopic quantum kinetic theory. 
We demonstrate that the spin transfer dynamics is qualitatively different
for components of the carrier spin parallel and perpendicular to the Mn magnetization.
From our quantum kinetic equations we have worked out 
the corresponding Markov limit which is equivalent to rate equations based on Fermi's
golden rule. The resulting equations resemble the widely used
Landau-Lifshitz-Gilbert-equations, but also describe genuine spin transfer due 
to quantum corrections. Although it is known that the Markovian rate description
works well for bulk systems when the initial Mn magnetization is zero, we find
large qualitative deviations from the full quantum kinetic theory for finite initial 
Mn magnetizations. These deviations mainly reflect corrections of higher than leading order 
in the interaction which are not accounted for in golden rule-type rates.
\end{abstract}
\maketitle
Diluted magnetic semiconductors (DMS) have been studied intensively in the past
decades, since they combine the versatility of semiconductors with 
the spin degree of freedom,  which promises future applications 
in spintronics\cite{Awschalom,Awschalom07,Spintronics,Zutic,MacDonald}.
The magnetic properties of DMS arise from the s/p-d exchange 
interaction\cite{DOM,Zutic,Wolff} between carriers and  
magnetic impurities, which 
typically consist of Mn ions acting as localized spin $\frac 52$ systems.
Especially for short timescales and high Mn doping concentrations the exchange interaction 
can dominate the spin dynamics \cite{Morandi2009,Wu09}.
The description of the resulting spin transfer dynamics in DMS  is usually based on  rate equations,
where the rates are  computed using Fermi's golden rule\cite{Koenig,Wu09}.
The standard derivation of the golden rule involves a Markov approximation\cite{Morandi2009,Cywinski} 
and is perturbative with respect to the exchange coupling constant.
In Ref.~\onlinecite{Semenov} a projection operator method was applied to
derive spin relaxation rates for DMS quantum wells. There, also a 
Markovian assumption as well as a perturbative argument were used.
Another approach to the description of the macroscopic
magnetization dynamics is the use of the phenomenological
Landau-Lifshitz-Gilbert equations\cite{Perakis2012,Morandi2011}.

Recently, starting from a Kondo-like interaction Hamiltonian 
a density matrix approach based on correlation expansion was 
developed\cite{Thurn:12} in order to describe the spin dynamics in the ultrafast
regime.  
Until now, this quantum kinetic theory (QKT) has only been applied to
the case of an initially zero Mn spin. There, it has been found that
in three dimensional systems, the time evolution of the carrier spin is
exponentially decreasing, where the decay rate coincides with its value according 
to Fermi's golden rule\cite{Thurn:13_1}. The latter was shown by performing the
Markov limit (ML) of the QKT using only terms in second order of $\Jsd$. 
In lower dimensional systems, excitation conditions can be found, where 
significant differences between the ML and the  QKT become 
visible although the memory induced by the exchange interaction is
orders of magnitude shorter than the timescale for the evolution of the
carrier and Mn dynamics \cite{Thurn:13_1}. In particular, quantum kinetic effects are most pronounced
when suitably tuned oppositely circular polarized two-color laser pulses are used for the excitation\cite{Thurn:13_2}.

In this article, we study the spin dynamics of conduction band electrons in a bulk ZnMnSe semiconductor 
for the case of a non-zero initial Mn spin where electron spins can precess around the Mn magnetization. 
It turns out that the spin transfer dynamics that is superimposed to the precession is qualitatively different
for electron spins aligned parallel or perpendicular to the Mn magnetization.
Starting from our quantum kinetic equations we derive the corresponding Markov limit
for finite Mn magnetization. The resulting equations can be interpreted 
as modified  Landau-Lifshitz-Gilbert equations. Assuming Mn concentrations much larger than
the itinerant electron density analytical solutions of these Markovian equations are presented.
The resulting analytical expressions also exhibit a different dynamics for 
perpendicular and parallel
spin transfer, which, however, quantitatively and qualitatively disagrees with the prediction
of the full QKT. Here, the failure of the Markovian approach can be traced back to contributions
of higher than leading order in the exchange coupling constant.

Outline of the paper: 
In a first step, we briefly summarize the QKT \cite{Thurn:12} that was used as a basis
for our numerical calculation and introduce the model used in this paper. 
Then, we derive the Markov limit  of the QKT along the lines described in Ref.~\onlinecite{Thurn:13_1} for 
an initially zero Mn magnetization 
$\langle \mbf{S}\rangle$, but allow for a finite value of $\langle \mbf{S}\rangle$ 
and an arbitrary angle between the conduction band electron spin and the Mn spin. 
In a subsequent section we present numerical results of our QKT 
for the spin transfer dynamics of the parallel and perpendicular components
and compare them with the ML. 
The analytical solution of the ML equations in combination with a 
rearrangement of the contributions to our QKT allows for a clear 
physical interpretation of the
pertinent source terms. By selectively studying the impact of different source terms
we are able to demonstrate the importance of contributions of higher than leading order
in the coupling constant.

\section{Quantum kinetic equations}
In Ref.~\onlinecite{Thurn:12}, a quantum kinetic density matrix approach 
for the spin dynamics in Mn doped semiconductors 
was developed starting from the Hamiltonian:
\begin{align}
  H = H_{0} + H_{sd} + H_{pd} + H_{em},
\end{align}
where $H_{0}$ describes the single particle band energies, $H_{sd}$ accounts for
the exchange interaction between the $s$-type conduction band electrons and
the spins of the $d$-type electrons of the Mn dopands while $H_{pd}$ stands for the
interaction of the latter with $p$-type holes. Finally, $H_{em}$ comprises the 
dipole coupling to an external laser field. 
The exchange interactions $H_{sd} + H_{pd}$ as well as the random spatial distribution
of Mn atoms give rise to a hierarchy of higher order correlation functions.
In order to obtain a finite set of dynamical variables 
a specially adapted  correlation expansion has been 
worked out in Ref.~\onlinecite{Thurn:12}. 

Since the aim of the present paper is to investigate the spin transfer between conduction band electrons
and Mn dopands,  the model can be reduced to:
\begin{align}
  H = H_{0} + H_{sd}.
\end{align}
$H_{0}$ now accounts only for electrons in a single spin degenerate conduction band:
\begin{align}
  H_{0} = \sum_{l\k} E_{\k} c^\dagger_{l\k}c_{l\k},
\end{align}
where
$c^\dagger_{l\k}$ ($c_{l\k}$) are the creation (annihilation) operators 
of conduction band electrons with k-vektor $\k$ and spin index $l=1,2$.
For simplicity we shall assume parabolic bands $E_{\k}=\frac{\hbar^{2} k^2}{2m^*}$,
with an effective mass $m^{*}$.
The exchange interaction is given by \cite{Zener, semicond_semimet}:
\begin{align}
H_{sd}&=J_{sd}\sum_{Ii} { \hat {\mbf S}}_{I}\cdot  { \hat {\mbf s}}_i^e \delta({\mbf r}_i - {\mbf R}_I),
\end{align}
where $J_{sd}$ is the exchange constant and  ${ \hat{ \mbf S }}_{I}$ ($ {\hat{ \mbf s}}_i^e$) are operators
for the spin  of the Mn atom (conduction band electron) in units of $\hbar$ at the position
 ${\mbf R}_I$ (${\mbf r}_i$). % \cite{Cywinski, Szczytko:96}. 
As in Ref.~\onlinecite{Thurn:12} we assume an on average
spatially homogeneous distribution of Mn positions  ${\mbf R}_I$.

According to the analysis in Ref.~\onlinecite{Thurn:12}
the relevant dynamical variables for this reduced model are:
\begin{subequations}
\begin{align}
\Cstd&=\langle c^\dagger_{l_1\k_1}c_{l_2\k_1}\rangle, \\
\Mstd&=\langle \hat{P}^I_{n_1n_2}\rangle, \\
\Kstd&=\delta \langle c^\dagger_{l_1\k_1}c_{l_2\k_2}\hat{P}^I_{n_1n_2}
e^{i(\k_2-\k_1)\mbf R_I}\rangle, \\
\Cqstd&=\delta \langle c^\dagger_{l_1\k_1}c_{l_2\k_2}e^{i(\k_2-\k_1)\mbf R_I}\rangle, 
\end{align}
\label{eq:variables}
\end{subequations}
where $\hat{P}^I_{n_1n_2} :=  |I,n_{1}\rangle \langle I,n_{2}|$ 
describes the spin state of the I-th Mn ion
($n=-\frac 52,\dots, \frac 52$).
The expectation value represented by the  brackets involves a quantum mechanical average as well as
the  disorder average over the randomly distributed Mn positions.
$\Cstd$ and $\Mstd$ are the electron and Mn density matrices. 
$\Kstd$ and $\Cqstd$ are the correlated parts of the corresponding density matrices, i.e.,
in these quantities all parts that can be factorized into products of lower order
correlations functions are subtracted from the expectation values.
The explicit but lengthy definitions of $\Kstd$ and $\Cqstd$ can be found
in Ref.~\onlinecite{Thurn:12}.

It turns out that the resulting equations of motion can be simplified by
introducing the following new correlation functions:
\begin{align}
\Qstd&:=\Kstd+\Mstd\Cqstd.
\label{eq:defQ}
\end{align}
Rewriting the equations of motion from  Ref.~\onlinecite{Thurn:12}
in terms of these functions we obtain:
\begin{widetext}
\begin{subequations}
\begin{align}
-i\hbar &\ddt \Mstd=\Jsd\frac 1V\sum_\k\sum_{nll'}\mbf s_{ll'}\bigg[\C{}{'}{}\big(
\mbf S_{nn_1}\M{}{_2}-\mbf S_{n_2n}\M{_1}{}\big)+
\frac 1V\sum_{\k'}\big(\mbf S_{nn_1}\Q{}{'}{}{_2}{}{'}-\mbf S_{n_2n}\Q{}{'}{_1}{}{}{'}\big)
\bigg],\\
-i\hbar &\ddt \Cstd=\Jsd\nMn \sum_{nn'l}\mbf S_{nn'}\bigg[\M{}{'}\big(
\mbf s_{ll_1}\C{}{_2}{_1}-\mbf s_{l_2l}\C{_1}{}{_1}\big)+
\frac 1V\sum_{\k}\big(\mbf s_{ll_1}\Q{}{_2}{}{'}{}{_1}-\mbf s_{l_2l}\Q{_1}{}{}{'}{_1}{}\big)
\bigg],\label{eq:eomddtC}\\
\label{eq:eomddtQ}
\bigg(&-i\hbar\ddt+E_{\k_2}-E_{\k_1}\bigg)\Qstd= \bstd^I+\bstd^{II}+\bstd^{III},
\allowdisplaybreaks\\
\intertext{with source terms}\nn
\label{eq:eomQb1}
&\bstd^I=
\underbrace{
\Jsd \sum_{nl}\bigg\{\mbf S_{nn_1}\mbf s_{ll_1}\C{}{_2}{_2}\M{}{_2}-\mbf S_{n_2n}\mbf s_{l_2l}
\C{_1}{}{_1}\M{_1}{}\bigg\}}_{=:\bstd^{I.1}}\underbrace{
-\Jsd\sum_{nll'}\mbf s_{ll'}\C{}{_2}{_2}\C{_1}{'}{_1}\big(\mbf S_{nn_1}\M{}{_2}
-\mbf S_{n_2n}\M{_1}{}\big)}_{=:\bstd^{I.2}},
\allowdisplaybreaks\\
\label{eq:eomQb2}
&\bstd^{II}=\underbrace{
\Jsd\sum_{nn'l}\mbf S_{nn'}\M{}{'}\nMn\big(\mbf s_{ll_1}\Q{}{_2}{_1}{_2}{_1}{_2}
-\mbf s_{l_2l}\Q{_1}{}{_1}{_2}{_1}{_2}\big)}_{\bstd^{II.1}}+\underbrace{
\Jsd \sum_{nll'}\mbf s_{ll'}\frac 1V\sum_\k \C{}{'}{}
\big(\mbf S_{nn_1}\Q{_1}{_2}{}{_2}{_1}{_2}-\mbf S_{n_2n}\Q{_1}{_2}{_1}{}{_1}{_2}\big)}_
{\bstd^{II.2}},
\allowdisplaybreaks\\
&\bstd^{III}=\underbrace{
\Jsd\sum_{nl}\bigg\{\frac 1V\sum_\k\big[ \mbf S_{nn_1}\mbf s_{ll_1}\Q{}{_2}{}{_2}{}{_2}
-\mbf S_{n_2n}\mbf s_{l_2l}\Q{_1}{}{_1}{}{_1}{}\big]\bigg\}}_{\bstd^{III.1}}\nn
&\qquad\qquad
\underbrace{-\Jsd\sum_{nll'}\mbf s_{ll'}\bigg\{\frac 1V\sum_\k \C{_1}{'}{_1}\big[\mbf S_{nn_1}
\Q{}{_2}{}{_2}{}{_2}-\mbf S_{n_2n}\Q{}{_2}{_1}{}{}{_2}\big]+\frac 1V\sum_\k
\C{}{_2}{_2}\big[\mbf S_{nn_1}\Q{_1}{'}{}{_2}{_1}{}-\mbf S_{n_2n}\Q{_1}{'}{_1}{}{_1}{}\big]\bigg\}}_{\bstd^{III.2}},
\end{align}
\label{eq:eomQ}
\end{subequations}
\end{widetext}
where $\mbf S_{n_1n_2}$ and $\mbf s^{e}_{l_{1}l_{2}}$ are the Mn and electron spin matrices, $V$ is the Volume 
of the DMS and $n_{Mn}=\frac{N_{Mn}}{V}$ is the density of the Mn ions.
We have subdivided the sources on the r.h.s. of Eq.~\ref{eq:eomddtQ}
for later reference. The physical meaning of these terms and their 
respective importance will be discussed later.

In order to study the dynamics of the spin transfer we consider initial conditions
where the electrons are initially spin polarized and the Mn magnetization corresponds 
to a thermal distribution while the correlations $\Qstd$ are assumed to be zero.
This is a situation typical for a system, 
immediately after an ultrafast optical excitation has induced a finite electron spin polarization.

\section{Markov limit}
It turns out to be instructive to derive the Markov limit of our QKT,
first of all, because this greatly simplifies the theory as the higher order correlation functions
are formally eliminated in favor of the variables of most interest, i.e., the
electronic densities and spins. Furthermore, the Markov limit provides a relevant
reference for our QKT. In particular for bulk systems it has been found previously \cite{Thurn:13_1} that
the memory of the exchange interaction is short and therefore it is tempting to think that
the Markovian equations should yield valid results in our case.

In order to be able to work out the Markov limit starting from Eqs.~(\ref{eq:eomQ}),
we follow the procedure that in Ref.~\onlinecite{Thurn:13_1}  led to rates in accordance with Fermi's
golden rule and neglect in a first step the source terms 
of higher than leading order in the exchange coupling $\Jsd$.
Due to the initial condition $\Qstd=0$ the correlations
$\Qstd$ are of first order in $\Jsd$ and thus we see from Eqs.~(\ref{eq:eomQ})
that $\bstd^{II}$ and $\bstd^{III}$ are of second order in $\Jsd$ and yield
third order contributions to the electron spin dynamics.
Thus, we keep in Eq.~(\ref{eq:eomddtQ}) only the first order term $\bstd^{I}$. 
This allows us to formally integrate the correlations:
\begin{align}
\Qstd(t)&=\frac i\hbar \int\limits_0^t dt' e^{i(\omega_{\k_2}-\omega_{\k_1})(t'-t)}
\bstd^{I}(t'),
\label{eq:integralgl}
\end{align}
with frequency $\omega_{\k}=\frac{E_{\k}}\hbar=\frac{\hbar k^2}{2m^*}$. 
Substituting Eq.~(\ref{eq:integralgl}) back into the equations for $\Cstd$
and $\Mstd$ we have to perform a $k$-summation which due to interference resulting 
from the $k$-dependent  phases $e^{i(\omega_{\k_2}-\omega_{\k_1})(t'-t)}$
leads to a finite memory.
The Markov limit is established by  assuming that the sources $\bstd^{I}$ change on
a much slower timescale than the memory and can therefore be drawn out of the integral.
The memory has been found to decay  on a fs-timescale while
the spin dynamics evolves 
on a ps-timescale \cite{Thurn:13_1}.
Therefore,  the lower limit of 
the integral can be extended to $-\infty$ resulting in the following
approximation for the correlations:
\begin{align}
&\Qstd(t)  \approx\frac i\hbar \bstd^{I}(t) \int\limits_{-\infty}^0 dt''
e^{i(\omega_{\k_2}-\omega_{\k_1})t''}\nn&=
\frac i\hbar \bstd^{I}(t)\left( \pi \delta(\omega_{\k_2}-\omega_{\k_1})
-\mathcal{P} \frac i{\omega_{\k_2}-\omega_{\k_1}} \right),
\label{eq:markov_app}
\end{align}
where $\mathcal{P}$ denotes the Cauchy principal value.

Starting from Eq.~(\ref{eq:eomddtC}) for the electron density 
$\Cstd$  we can set up an equation of motion
for the average electron spin 
$\langle \mbf s_{\k_1}\rangle=\sum_{l_1l_2}\mbf s^e_{l_1l_2}\Cstd$
in the state with k-vector $\k_1$.
Feeding back the correlations $\Qstd$ from Eq.~(\ref{eq:markov_app})
into these equations we finally obtain:
\begin{widetext}
\begin{align}
\ddt \langle \mbf s_{\k_1}\rangle&=\frac{\Jsd n_{Mn}}\hbar
(\langle \mbf S\rangle\times\langle \mbf s_{\k_1}\rangle)+
\frac{\Jsd^2n_{Mn}}{\hbar^2V} \sum_\k \Bigg\{\frac 12
\mathcal{P}\frac{n_\k-1}{\omega_{\k_1}-\omega_\k}
\big(\langle \mbf S\rangle \times \langle \mbf s_{\k_1}\rangle\big)
\nn&+\pi \delta(\omega_{\k_1}-\omega_{\k})
\bigg[
\langle \mbf S\rangle\frac{4\langle \mbf s_{\k_1}\rangle^2-n_{\k_1}^2+2n_{\k_1}}4+
\big(\langle\mbf s_{\k}\rangle\times\big(\langle\mbf s_{\k_1}\rangle\times\langle \mbf S\rangle
\big)\big)
+\frac{\langle \mbf S\times (\mbf S \times \langle
\mbf s_{\k}\rangle)\rangle+\langle (\langle\mbf s_{\k}
\rangle\times \mbf S)\times \mbf S\rangle}2
\bigg]
\Bigg\}.
\label{eq:markov}
\end{align}
\end{widetext}
Applying the same procedure to the electron occupations $n_{\k_1}=\sum_{l}\C{}{}{_1}$  
at a given k-vector $\k_1$ we find that on this level of theory $n_{\k_1}$ is time independent.
It should be noted, that in the full QKT this is not the case. Instead it was shown 
in Refs.~\onlinecite{Thurn:12,Thurn:13_1,Thurn:13_2} that  redistributions in $k$-space take place which are
responsible for a number of features of the magnetization dynamics 
that are not expected in the Markovian theory.

The different terms in equation (\ref{eq:markov}) can easily be interpreted.
The first term describes the precession of the electron spin in an effective 
magnetic field due to the Mn magnetization $\langle \mbf S\rangle$, which is also the result
of a mean-field calculation\cite{Thurn:12}. The second term represents a renormalization
of the precession frequency that depends on the density of states and therefore on
the dimensionality of the system as well as the k-vektor, which can possibly lead to 
dephasing of the electron spin.\\
The magnitude of the renormalization for a bulk semiconductor can be estimated 
in the continuum limit by 
approximating the Brillouin zone (BZ) as a sphere with radius $k_{BZ}$
and assuming a parabolic band structure
as follows:
\begin{align}
\Delta\omega_M&=\omega_M^0 \frac{\Jsd}{\hbar (2\pi)^2}\frac{2m^*}{\hbar}
\underbrace{\int\limits_0^{k_{BZ}}dk \frac{k^2}{k^2-k_1^2}(1-n_{\k})}_{
\approx k_{BZ}},
\label{eq:renorm3d}
\end{align}
where $\omega_M^0=\frac{\Jsd n_{Mn}}\hbar |\langle \mbf S\rangle|$ is the mean-field precession 
frequency. The order of magnitude of the integral on the r.h.s. of 
Eq.~(\ref{eq:renorm3d}) can be determined by noting 
that the optically excited carriers occupy only very few states near the center of the BZ 
and therefore for the most part of the BZ $n_{\k}\approx 0$ holds which also implies
$\frac {k_1}{k}\ll 1$ for the occupied states. Approximating $n_{\k}\approx 0$ and
 $\frac {k_1}{k}\approx 0$ the integral yields the value $k_{BZ}$.
For the parameters used in our study (see below) the 
renormalization is  estimated in this way 
to be of the order of  $\approx 1\%$ of the  mean-field
precession frequency\footnote{For lower dimensional systems this crude approximation 
leads to a divergence of the frequency renormalization 
at $k\rightarrow k_1$. This fact supports the
findings of Refs.~\onlinecite{Thurn:13_1,Thurn:13_2} that the Markov limit is not a 
good approximation in systems with dimensions lower than 3.}.\\
The third term in Eq.~(\ref{eq:markov}), which is proportional to the Mn spin,  
describes a transfer of spin from the Mn to the electron system.
The prefactor $\frac{4\langle \mbf s_{\k_1}\rangle^2-n_{\k_1}^2+2n_{\k_1}}4$ 
is zero for $n_{\k_1}\in\{0,2\}$. For $n_{\k_{1}}=0$ no transfer can occur because
there are no electrons that can exchange their spins with the Mn atoms; for
$n_{\k_{1}}=2$ the transfer vanishes due to Pauli blocking.

The term proportional 
to $\langle\mbf s_{\k}\rangle\times\big(\langle\mbf s_{\k_1}\rangle\times\langle \mbf S\rangle\big)$
has the form of the relaxation term of a Landau-Lifshitz-Gilbert (LLG) equation
and describes the tendency of a spin  in a given effective magnetic field to align
along the direction of the field. Unlike in the LLG equation, here, the prefactor is determined
by the parameters of the microscopic model and is not a phenomenological
fitting parameter.

The last term in Eq.~(\ref{eq:markov}) resembles a relaxation term that would be expected 
in the LLG equation for the Mn magnetization $\langle \mbf S \rangle$. 
Here, it arises in the equation for the electron spin reflecting the 
conservation of total spin which is a feature also of the full QKT \cite{Thurn:12}.
However, there is a crucial difference between the last term in Eq.~(\ref{eq:markov})
and the LLG relaxation term for the Mn magnetization:
while the cross-products in the LLG equation involve classical vectors,
we are dealing here with vector operators.  
Here, the expectation value has to be taken 
after constructing the cross product in a symmetrized form.  
The physical consequences of this difference become most obvious
by rewriting the last term in Eq.~(\ref{eq:markov})
as follows:
\begin{align}
&\frac{\langle \mbf S\times (\mbf S \times \langle
\mbf s_{\k}\rangle)\rangle+\langle (\langle\mbf s_{\k}
\rangle\times \mbf S)\times \mbf S\rangle}2=\nn&
-\big(\langle S^2\rangle -\langle {S^{\|}}^2\rangle \big)\langle \mbf s_{\k}^\|\rangle
-\frac 12\big(\langle S^2\rangle +  \langle {S^{\|}}^2\rangle\big)
\langle \mbf s_{\k}^\perp\rangle,
\label{eq:damping}
\end{align}
where $\langle \mbf s_{\k}^\|\rangle$ and $\langle \mbf s_{\k}^\perp\rangle$ 
describe the electron spin of the states with k-vector $\k$ in the direction 
parallel and perpendicular to the Mn spin vector $\langle \mbf S \rangle$
and $S^{\|}= \mbf S\cdot \frac{\langle \mbf S \rangle}{|\langle \mbf S \rangle|} $.

It is seen from Eq.~(\ref{eq:damping}) that even when the electron spin is
aligned parallel to the Mn spin, a spin transfer can occur, and
it was already noted in Ref.~\onlinecite{Thurn:13_1}
 that the corresponding parallel spin transfer rate 
coincides with the result of Fermi's golden rule. In contrast,  the
corresponding term in the standard  LLG equation would be zero. 
This transfer is
enabled because  the factor $\langle S^2\rangle -\langle {S^{\|}}^2\rangle$ is
non-zero as quantum mechanically  the maximal value of $\langle {S^{\|}}^2\rangle$
is $\hbar^{2} S^{2}$, while  $\langle S^2\rangle=\hbar^{2}S(S+1)$ which reflects the 
uncertainty between the respective spin components. 
For classical vectors, as considered in the standard LLG equation, this factor would always be zero. 
Furthermore, in general the contribution in Eq.~(\ref{eq:damping}) 
is different for the parallel and perpendicular components of the electron spin.
It is noteworthy that if the Mn spin had been represented by a pseudo-spin $\frac{1}{2}$,
 this feature would be lost as then independent of the Mn spin configuration 
we find $\langle S^2\rangle=\frac{3}{4}$ and $\langle {S^{\|}}^2\rangle=\frac{1}{4}$
resulting in the same prefactors for $\langle \mbf s_{\k_1}^\|\rangle$
and $\langle \mbf s_{\k_1}^\perp\rangle$  in Eq.~(\ref{eq:damping}).

In order to use Eq.~(\ref{eq:markov}) in practical calculations we have to know
the values of the average Mn spin $\langle \mbf S\rangle$ and according to Eq.~(\ref{eq:damping})
the second moment $\langle{S^{\|}}^2\rangle$ which appear on the r.h.s.
of Eq.~(\ref{eq:markov}).
The average Mn spin can be calculated from the knowledge of the
electron spin and the initial total spin by using the total spin 
conservation\cite{Thurn:12}. Setting up an equation of motion for the second moment is
cumbersome and not necessary  for the cases that we shall discuss in this paper
where it is assumed that the number of Mn ions by far exceeds the number of photo induced 
electrons ($\NMn\gg N_e$).
In this case, the change of the average Mn spin as well as its
second moment can be neglected and thus the second moment 
essentially coincides with its initial thermal value.
Furthermore, for nearly constant Mn magnetization,
the equations of motion for electron states with different 
energies $\hbar\omega_{\k}$ are decoupled
in the Markov limit due to the delta-distribution in Eq.~(\ref{eq:markov})
and the fact that $n_\k$ remains constant
which allows using the initial occupation for
the evaluation of the frequency renormalization.

\begin{figure*}[t!]
\centering
\includegraphics{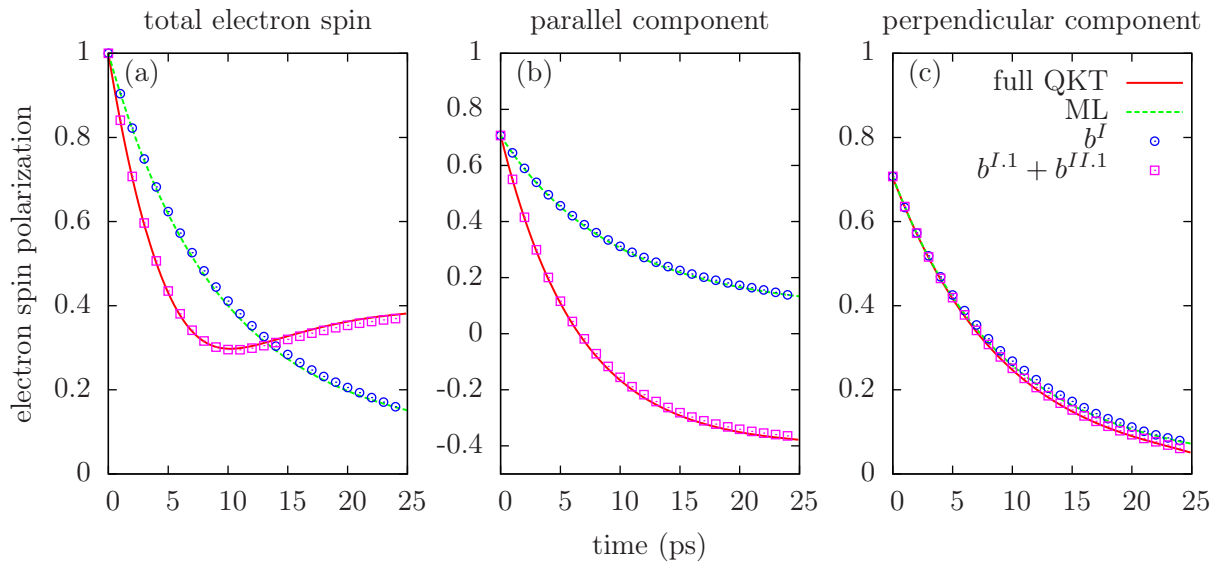}
\caption{(Color online) 
Time evolution of the total electron spin polarization (a) and its components parallel (b)
and perpendicular (c) to the Mn spin assuming the electrons to be initially 
spin polarized along a direction at an angle of $45^{\circ}$ relative to
the Mn magnetization.
The solid red line describes the spin
dynamics according to the full quantum kinetic theory, the dashed green line shows its
Markov limit (analytic solutions, cf. appendix~\ref{ap:riccati}). Blue circles
 and purple squares correspond
to approximate quantum kinetic calculations where only a subset of source terms for the
correlations (as indicated in the key of the figure) has been accounted for.}
\label{fig:comp}
\end{figure*}

The decoupling of the equations of motion in the Markov limit 
enables us to find analytical solutions 
for Eq.~(\ref{eq:markov}). 
To this end we split the electron spin into its components
parallel and perpendicular to the Mn spin 
according to:
%\begin{widetext}
\begin{align}
\langle \mbf s_{\k_1}\rangle&=s^\|_{\k_1}\frac{\langle \mbf S\rangle}S 
+s^\perp_{\k_1}\bigg(\sin (\omega_M t)\frac{\langle \mbf S\rangle\times
\langle \mbf s_{\k_1}(0)\rangle}{|\langle \mbf S\rangle\times
\langle \mbf s_{\k_1}(0)\rangle|}\nn&+
\cos (\omega_M t)\frac{(\langle \mbf S\rangle\times
\langle \mbf s_{\k_1}(0)\rangle)\times \langle \mbf S\rangle}
{|(\langle \mbf S\rangle\times
\langle \mbf s_{\k_1}(0)\rangle)\times \langle \mbf S\rangle|}\bigg),
\label{eq:e-spin}
\end{align}
where $\omega_{M}$ accounts for the precession of the perpendicular component
that results from Eq.~(\ref{eq:markov}).
With this decomposition, Eq.~(\ref{eq:markov}) can be rewritten as:
\begin{subequations}
\begin{align}
\label{eq:ricc_par}
\ddt s^\|_{\k_1}&=\gamma_{\k_1} S
(\mbf s^\|_{\k_1})^2 +\gamma_{\k_1} S \frac{n_{\k_1}(2-n_{\k_1})}4\nn&
-\gamma_{\k_1}\big(\langle S^2\rangle-\langle{S^\|}^2\rangle\big)s^\|_{\k_1},\\
\label{eq:ricc_per}
\ddt s_{\k_{1}}^\perp&=\gamma_{\k_1} s^\|_{\k_1}s^\perp_{\k_1} S
-\frac 12\gamma_{\k_1}\big(\langle S^2\rangle+\langle{S^\|}^2\rangle\big)s^\perp_{\k_1},
\end{align}
\label{eq:ricc}
\end{subequations}
with
\begin{subequations}
\begin{align}
\label{eq:gamma}
\gamma_{\k_1}&=\frac{\Jsd^2\nMn}{\hbar^2V}\pi\sum_{\k}\delta(\omega_{\k_1}
-\omega_{\k}),\\
\omega_M&=\frac{\Jsd\nMn}{\hbar}S\bigg(1+\frac 12\frac{\Jsd}{\hbar V}
\sum_\k\mathcal{P}\frac{n_\k-1}{\omega_{\k_1}-\omega_{\k}}\bigg),
\label{eq:ricc_freq}
\end{align}
\end{subequations}
and $S=|\langle \mbf S\rangle|$. Eq.~(\ref{eq:ricc_par}) is a Riccati 
differential equation with constant coefficients which can be solved 
analytically. Its 
solutions can then be fed back into Eq.~(\ref{eq:ricc_per})
for the perpendicular electron spin. 
The explicit solutions are listed in appendix~\ref{ap:riccati}.

It is noteworthy that by a rescaling of the time axis according to $\tau:= \gamma_{\k_1} t$
all material parameters can be eliminated from Eqs.~(\ref{eq:ricc}) for 
the moduli $s^\|_{\k_1}$ and $s^\perp_{\k_1}$. 
Therefore, with this choice of time units
and  given initial conditions we obtain 
the same universal solution for all material parameters.
Reinserting the solutions for $s^\|_{\k_1}$ and $s^\perp_{\k_1}$ into
Eq.~(\ref{eq:e-spin}) and choosing again $1/\gamma_{\k_1}$ as the unit
of time, we conclude that for given initial conditions
the time trace of the electron spin $\langle \mbf s_{\k_1}\rangle$
is affected by the material parameters only via the ratio
$\omega_M/\gamma_{\k_1}$.

\section{Numerical results}
\label{numerics}
The quantum kinetic equations of motion (\ref{eq:eomQ}) have been solved numerically 
and compared with their Markov limit (\ref{eq:markov}) for different  initial conditions 
in a three dimensional bulk DMS. 
The initial electron distribution over the single particle energies $E_{\k}$ is taken to 
be Gaussian with its center at $E_{k=0}$ and a standard deviation of $\sigma=3\t{ meV}$
while the initial magnitude of the Mn spin is  set to $\frac 12 \hbar$ (i.e., $20\%$ of its maximal value). 
The material
parameters used were the same as in Ref.~\onlinecite{Thurn:13_1} for Zn$_{0.93}$Mn$_{0.07}$Se %bulk semiconductor
with $\Jsd=12\t{ meVnm}^3$ and $m_e=0.21 m_0$. 

First, we shall discuss results where at the beginning of the simulation  the electron spins 
are assumed to be totally polarized in a direction with an angle of $45^{\circ}$ with respect to the 
Mn magnetization vector. 
Displayed in Fig.~\ref{fig:comp} is the corresponding time evolution of 
the electron spin; part (a) shows the
total electron spin, while in parts (b) and (c) 
the components parallel and perpendicular to the
Mn magnetization are plotted, respectively. 
The full quantum kinetic results are plotted as solid red lines whereas 
curves derived from the analytical solutions of the Markov limit
equations are depicted as dashed green lines.

As seen from Fig.~\ref{fig:comp} (a), the  dynamics predicted by the full theory
is qualitatively different from the Markovian result. 
On a short time scale (for our parameters $t < 5$ ps), 
the electron spin decays much faster for the full solution 
than in the Markov limit. Subsequently, the quantum kinetic curve
exhibits a non-monotonic time dependence and 
the electron spin eventually approaches a finite value.
In contrast, in the Markov limit, we find a monotonic, 
almost exponential decay for all times. 
From the explicit analytical expression (cf. appendix~\ref{ap:riccati}) 
it is seen that the long time limit of the electron spin 
in the Markov limit is zero.

The origin of the non-monotonic  behavior can be understood by splitting
the total electron spin into its components parallel [Fig.~\ref{fig:comp} (b)]
and perpendicular [Fig.~\ref{fig:comp} (c)] to the Mn spin.
Both spin components decrease 
almost exponentially in the ML as well as in the full QKT.
The time evolution of the perpendicular spin component essentially yields the same results
for the full quantum kinetic calculation and the Markov limit. 
In the full QKT, however, the parallel spin component changes its sign and converges to
a finite negative value, whereas both spin components in the ML and the perpendicular 
spin component of the QKT drop to zero.
When the parallel spin component in the full QKT crosses the zero line, its modulus has a minimum
which leads to a minimum in the total spin.

The obvious discrepancy between the different levels of theory with regard to the 
dynamics of the parallel spin component does not arise
due to the assumption of a short memory in the ML. 
This can be seen from calculations, where only the source terms $\bstd^{I}$, 
i.e., the terms used to derive the ML in the first place, have been taken into 
account but the
finite memory expressed by the retardations in Eq.~\ref{eq:integralgl}  are still 
kept [blue circles in Fig.~\ref{fig:comp}].
The resulting curves almost coincide with the Markovian calculation. 
The main difference between the full QKT and the ML is due to the source term 
$\bstd^{II.1}$,
which is demonstrated by simulations that incorporate only 
$\bstd^{I.1}$ and $\bstd^{II.1}$ [purple squares in Fig.~\ref{fig:comp}]. 
The results of these calculations
agree very well with the predictions of the full theory, suggesting that all
other source terms are of minor importance, at least for the parameters used
here.
It should be noted, that especially the term $\bstd^{II.1}$,
like $\bstd^{II.2}$ and $\bstd^{III}$, gives
contributions to the reduced electron 
density matrices in the order of $\mathcal{O}(\Jsd^{3})$ while the leading order contributions of the
correlations are of $\mathcal{O}(\Jsd^{2})$. Thus, our results imply that a proper description of the 
coupled electron and Mn spin dynamics requires a treatment beyond perturbation theory. 
\begin{figure}[t!]
\centering
\includegraphics{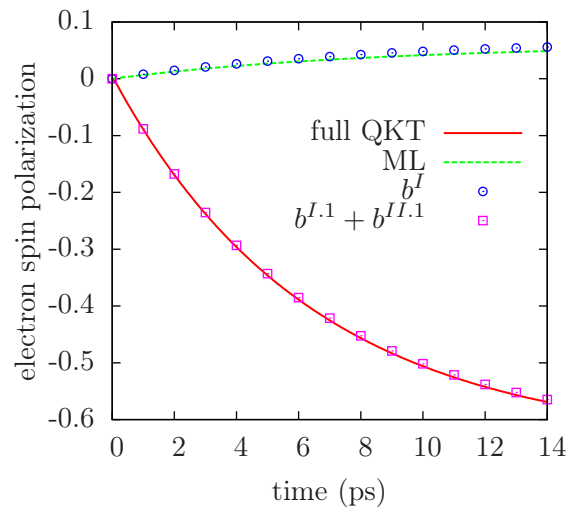}
\caption{(Color online) 
Dynamics of the electron spin polarization for initially unpolarized electron spins.
Line styles and symbols have the same meaning as in Fig.~\ref{fig:comp}.
}
\label{fig:s0}
\end{figure}

The effect of these higher order contributions on the 
dynamics is particularly dramatic in the case
of initially unpolarized electron spins. Corresponding results are displayed in Fig.~\ref{fig:s0}. Here,
even the sign of the spin polarization is opposite for the QKT and ML 
calculations. Furthermore, also the predictions concerning 
the magnitude of the spin polarization deviate significantly.

\section{Interpretation of the source terms}
By the numerical analysis in the last section, we were able to trace back the difference between 
the full quantum kinetic theory and its Markov limit to a few selected source terms for the 
correlations in Eqs.~(\ref{eq:eomQ}).
In this section, we shall give a physical interpretation to the individual source terms which will
enable us to understand what determines their relative importance.

First of all, $\bstd^{I.1}$ is the most important source term, because it starts the correlation 
dynamics, i.e., without these sources the correlations would stay zero for all times.
In the Markov limit, $\bstd^{I.1}$ yields a Landau-Lifshitz-Gilbert-like 
damping term described in 
Eq.~(\ref{eq:damping}) and a spin transfer term proportional to 
the Mn spin $\langle \mbf S\rangle$.  
$\bstd^{I.2}$ provides corrections for Pauli blocking to the transfer term and yields 
another LLG-like damping term, where the electron spin appears twice in the double cross product
(cf. Eq.~(\ref{eq:markov})).  
As seen above, the quantum kinetic $\bstd^{I}$ contributions act 
similarly to their Markov limit counterparts. 
The dominant role  of these terms is further emphasized by the fact that they are the leading 
terms in a perturbative treatment with respect to the exchange coupling constant $\Jsd$.

In order to understand the meaning of the $\bstd^{II}$ terms, it is instructive to reformulate the
equations of motion of the QKT by introducing new correlation functions according to:
\begin{align}
Q^{\alpha \k_2}_{\beta \k_1}:=\sum_{\stackrel{l_1l_2}{n_1n_2}}S^\beta_{n_1n_2}s^\alpha_{l_1l_2}
\Qstd,
\label{eq:defQGeom}
\end{align}  
which are summed over the electron band and Mn state indices.
Here, we use the conventions $\alpha=0\dots 3$ with $s^0_{l_1l_2}=\delta_{l_1l_2}$ and $\beta=1\dots 3.$ 
From Eq.~(\ref{eq:eomddtQ}), we obtain the following equations of motion for the summed correlations:
\begin{subequations}
\begin{align}
\ddt Q^{0\k_2}_{\beta\k_1}=&-i(\omega_{\k_2}-\omega_{\k_1})Q^{0\k_2}_{\beta\k_1}
+{b^{0\k_2}_{\beta\k_1}}^{\t{Res}}\\
\ddt  Q^{{\alpha}\k_2}_{\beta\k_1}=&-i(\omega_{\k_2}-\omega_{\k_1})
Q^{{\alpha}\k_2}_{\beta\k_1}+{b^{{\alpha}\k_2}_{\beta\k_1}}^{\t{Res}}\nn&
+\sum_{\kappa\lambda}\epsilon_{\alpha \kappa\lambda }\omega_M^{\kappa}Q^{\lambda\k_2}_{\beta\k_1}
+\sum_{\kappa\lambda}\epsilon_{\beta\kappa\lambda} \omega_{E}^{\kappa}Q^{\alpha\k_{2}}_{\lambda\k_{1}},
\label{eq:corr_geom_c}
\end{align}
\label{eq:corr_geom}
\end{subequations}
where
\begin{subequations}
\begin{align}
\omega^{\alpha}_M=&\frac \Jsd\hbar \nMn \langle  S^{\alpha}\rangle, \\
\omega^{\alpha}_E=&\frac \Jsd\hbar \frac 1V\sum_{\k} \langle  s_{\k}^{\alpha}\rangle, \\
{b^{\alpha\k_2}_{\beta\k_1}}^{\t{Res}}=&\sum_{\stackrel{l_1l_2}{n_1n_2}}
S^\beta_{n_1n_2}s^\alpha_{l_1l_2}\big[\bstd^{I}+\bstd^{III}\big]
\end{align}
\end{subequations} 
and $\epsilon_{\alpha\beta\gamma}$ is the Levi-Civita symbol.
%\begin{align}
%({\mbf b^{\phantom{\alpha}\k_2}_{\beta\k_1}}^{\t{Res}})_\alpha={b^{\alpha\k_2}_{\beta\k_1}}^{\t{Res}}=
%\sum\limits_{\stackrel{l_1l_2}{n_1n_2}}S^\beta_{n_1n_2}s^\alpha_{l_1l_2}\bstd^{\t{Res}},
%\end{align}
%with 
We note in passing that the residual sources 
${b^{\alpha\k_2}_{\beta\k_1}}^{\t{Res}}$
contain a term resulting 
from $\bstd^{III.1}$ which cannot be expressed by the summed 
correlations. Thus, Eqs.~(\ref{eq:corr_geom})
are numerically advantageous only if $\bstd^{III.1}$ is disregarded.
The point here is that the two terms in Eq.~(\ref{eq:corr_geom_c}) 
originating from $\bstd^{II.1}$ and $\bstd^{II.2}$
both involve the Levi-Civita symbol and can therefore be interpreted as describing precessions. This can be made more explicit,
e.g., by introducing a vector with components $\alpha$ according to
\begin{align}
(\mbf Q^{\phantom{\alpha}\k_2}_{\beta\k_1})_\alpha=Q^{\alpha\k_2}_{\beta\k_1}.
\end{align}
Then, the first of these terms, which stems from $\bstd^{II.1}$, can be written as a cross product:
\begin{align}
\boldsymbol{\omega}_{M}\times \mbf{Q}^{\phantom{\alpha}\k_{2}}_{\beta\k_{1}}
\end{align}
indicating a precession of the vector $\mbf{Q}^{\phantom{\alpha}\k_{2}}_{\beta\k_{1}}$ around the direction 
$\boldsymbol{\omega}_{M}$ of the Mn magnetization with the same frequency as the mean field precession of the 
electron spin. Likewise, the term originating from $\bstd^{II.2}$ has a similar structure. It can also be written as
a cross product
\begin{align}
\boldsymbol{\omega}_{E}\times \mbf{Q}^{\alpha\k_{2}}_{{\phantom{\beta}}\k_{1}},
\end{align}
 where now the index $\beta$ is associated with the components of a vector 
$\mbf{Q}^{\alpha\k_{2}}_{{\phantom{\beta}}\k_{1}}$ formed from the correlations according to 
\begin{align}
(\mbf Q^{\alpha\k_2}_{\phantom{\beta}\k_1})_\beta=Q^{\alpha\k_2}_{\beta\k_1},
\end{align}
 i.e., now we are dealing with a precession around the direction 
$\boldsymbol{\omega}_{E}$ of the electron spin.
Thus, not only the average spins of the electrons and Mn atoms exhibit a 
precession dynamics, but also their correlations, which is represented in 
the equations of motion by the $\bstd^{II}$ terms.

Finally, the physical meaning of the $\bstd^{III}$ source terms becomes clear by noting that their structure is
analogous to the structure of the $\bstd^{I}$ terms, where the products of electron and Mn density matrices 
are replaced by the corresponding unfactorized correlation functions.
Thus, the $\bstd^{III}$ sources provide the correlated parts of the $\bstd^{I}$ sources which represented a
Landau-Lifshitz-Gilbert-like dynamics including Pauli blocking.

Now that all source terms have been physically interpreted, let us come back to the question
of their relative importance in the case considered numerically in subsection \ref{numerics}.
As already noted, the sources $\bstd^{I}$  always play a pivotal role, since
no correlations would build up without these terms.  
The importance of the remaining terms depends on the physical situation.
Looking at the definition Eq.~(\ref{eq:eomQ}d-f) of the sources, it is seen that the terms
$\bstd^{X.2}$, with $X\in\{I,II,III\}$,  comprise similar factors as the corresponding 
contributions $\bstd^{X.1}$, except that the former contain an additional factor 
proportional to the electron density matrix $\Cstd$. From this observation we can conclude that 
the $\bstd^{X.2}$ sources should be less important than the $\bstd^{X.1}$ terms, if
the electron density is moderate, as it is the case here. 
A criterion for being in the low density limit is particularly easy to formulate for the $\bstd^{II}$ 
terms, since Eq.~(\ref{eq:eomQ}e) implies that $\bstd^{II.2}$ is negligible compared with $\bstd^{II.1}$ 
if $N_{Mn}\gg N_{e}$ which is fulfilled in our simulations. 
However, it is more challenging to give a condition for the 
negligibility of the $\bstd^{I.2}$ term, as it strongly depends on the 
electron distribution in k-space.

Finally, since the $\bstd^{III}$ sources have the same structure as the $\bstd^I$ term, 
except that the correlations $\Qstd$ take the place of the product $\Cstd\Mstd$, they will be of minor importance if
the relation $\frac{\Qstd}{\Cstd\Mstd}\ll 1$ is satisfied. 
The latter relation is expected to hold, when the conditions for the applicability of the correlation
expansion are fulfilled. The numerical results 
shown in Fig.~\ref{fig:comp} indicate that the $\bstd^{III}$ terms
provide insignificant quantitative corrections which confirms 
the consistency of the correlation expansion approach.

The fact that a source contains correlations is, however, not sufficient for concluding that it can be neglected compared
with the $\bstd^{I}$ terms, which do not involve correlations. In particular, the $\bstd^{II.1}$ term was shown to 
qualitatively modify the spin dynamics (cf. Figs.~\ref{fig:comp} and \ref{fig:s0}).
In view of our interpretation of the $\bstd^{II.1}$ term, this implies physically that accounting for the  
precession of the correlations around the Mn magnetization is essential for a correct description of the spin dynamics. 
This also explains why 
previous studies in Refs.~\onlinecite{Thurn:13_1}  and \onlinecite{Thurn:13_2} reported 
a negligible contribution from the $\bstd^{II.1}$ term, since there a situation was considered, where 
the average Mn spin was initially set to zero which suppresses the precession.

The features of the spin dynamics predicted in this article manifest themselves in the time evolution
of the spin polarization which is a quantity accessible experimentally, e.g., by 
time- and polarization-resolved photoluminescence or Faraday-/Kerr-rotation measurements\cite{SpinPhysSemi5}.
%For the quantitative description of experiments, however, other spin relaxation and dephasing mechanisms that have been 
%neglected in this treatment, in particular the D'yakonov Perel' mechanism\cite{DP}, could become important\cite{WuReview}
%(cf. Ref.~\onlinecite{Wu09} for a discussion of the importance of the different mechanisms in DMS).
%herefore, a particular interesting 
Favorable for the observation of such effects should be experiments 
measuring the time dependence of the spin polarization as well as the
its equilibrium value where
the angle between the Mn magnetization and the initial electron spin polarization induced by a circularly polarized laser beam
is varied. 
For our purposes, bulk materials are preferable compared with, e.g. quantum wells, since 
for heterostructures, the anisotropy with respect to growth axis 
as well as structure inversion asymmetry can play a role\cite{Rashba} which would make it hard to separate the angular 
dependence predicted by our theory from anisotropy effects.
Furthermore, II-VI DMS should be better suited for the proposed experiment than III-V DMS, since
they have the advantage of isoelectrical doping. In III-V materials,
the Bir-Aronov-Pikus interaction\cite{BAP} between electron and hole spins can  dominate the spin dynamics\cite{Wu09},
while for II-VI DMS with sufficiently high Mn doping the s-d-exchange interaction is typically the most important spin
relaxation mechanism\cite{WuReview}.

\section{Summary}
In this article, we have analyzed the spin dynamics of conduction band electrons in Mn doped
bulk DMS induced by the s-d-exchange interaction. 
In contrast to our previous studies \cite{Thurn:13_1,Thurn:13_2}, we now assume a non-zero Mn magnetization.
This naturally leads to a distinction between the electron spin dynamics of the components parallel and 
perpendicular to the Mn spin which introduces an anisotropy in the spin relaxation. 
Starting from a microscopic quantum kinetic theory based on correlation expansion 
we have derived the Markov limit yielding equations similar to the widely used phenomenological 
Landau-Lifshitz-Gilbert equations. Our derivation yields microscopic expressions for the parameters in
the Landau-Lifshitz-Gilbert equations and allows us to identify  some quantum corrections. 
The resulting rate equations  were solved analytically.
 
Numerical simulations within the quantum kinetic theory revealed that, while the dynamics of the perpendicular electron spin
component can be well described by the Markovian theory, the parallel component exhibits qualitative deviations 
between the full quantum kinetic and the corresponding Markovian results. 
The differences between both levels of theory manifest themselves in a non-monotonic temporal behavior of the 
total spin in the quantum kinetic theory as opposed to an almost exponential monotonic decay predicted by the Markovian
theory. Moreover, for certain excitation conditions, even the sign of the spin polarization differs between these levels
of theory.

A detailed analysis allowed us to assign a physical interpretation to all source terms for the correlations
and to understand their relative importance found in our numerical studies.
With the help of this analysis and our numerical results,
the deviations between the full quantum kinetic theory and its Markov limit
were traced back to the neglect of a precession dynamics of the 
correlations in the Markov theory. This precession is missing in the Markov limit
not because of the assumption of a short memory but 
due to the perturbative treatment that is implicit in this approach. 

\begin{acknowledgments}
We acknowledge the support by the Deutsche Forschungsgemeinschaft
through the Grant No. AX 17/9-1.
\end{acknowledgments}

\begin{appendix}
\section{Analytical solutions of the Markov equations}
\label{ap:riccati}
Eq.~(\ref{eq:ricc}a) is a Riccati differential equation
\begin{align}
\label{eq:app_ricc}
\ddt \spar=f \spar^2 -g \spar +h,
\end{align}
with $f=\gamma_{\k_1}S$, $g=\gamma_{\k_1}\big(\langle S^2\rangle -
\langle{S^\|}^2\rangle\big)$ and $h=\gamma_{\k_1}S\frac{n_{\k_1}(2-n_{\k_1})}4$.
For $f=0$, which is the case if $S=0$, the solution of Eq.~(\ref{eq:app_ricc})
is simply:
\begin{align}
\spar(t)=\big(\spar(0)-\frac hg\big)e^{-gt}+\frac hg.
\end{align}
For $f\neq 0$, the Riccati equation can be rewritten in terms of a linear
differential equation with eigenvalues:
\begin{align}
&\lambda_{1/2}=-\underbrace{\frac g2}_{=:\mu}\pm 
\underbrace{\sqrt{\frac {g^2}4-fh}}_{=:\nu}.
\end{align}
The solution of Eq.~(\ref{eq:app_ricc}) is then given by:
\begin{subequations}
\begin{align}
\spar(t)&=\frac \mu f -\frac \nu f\t{tanh}\left(\frac\varphi 2+\nu t\right)%\\
%	&=\frac{\mu-\nu}f+2\frac \nu f \frac 1{e^{\varphi+2\nu t}+1},
\end{align}
\label{eq:spar_loes}
\end{subequations}
where $\varphi$ is determined by the initial value of $\spar$.

Eq.~(\ref{eq:ricc}b) for the perpendicular spin component assumes the form:
\begin{align}
\ddt \sper&=\big(-\xi+f \spar\big)\sper,
\label{eq:sper_dgl}
\end{align}
where $\xi=\frac 12 \gamma_{\k_1}\left(\langle S^2\rangle +
\langle {S^\|}^2\rangle\right)$. Eq.~(\ref{eq:sper_dgl}) is solved by
\begin{align}
&\sper(t)=\sper(0)e^{-\xi t}\underbrace{e^{f\int\limits_0^t \spar(t')dt'}}_{
=:I}.
\end{align}
For $f=0$, $I=1$ and the perpendicular spin component decreases exponentially.
Inserting the solution for the parallel spin compontent from 
Eq.~(\ref{eq:spar_loes}) for non-zero $f$ yields: 
\begin{subequations}
\begin{align}
I&=e^{\mu t}\frac{\t{cosh}\left(\frac \varphi 2\right)}{
\t{cosh}\left(\frac \varphi 2+\nu t\right)} %\\
%&=e^{(\mu-\nu)t} \frac{e^{-\varphi}+1}{e^{-(\varphi+2\nu t)}+1}.
\end{align}
\end{subequations}
\end{appendix}
%\bibliography{zitate}

\begin{thebibliography}{24}%
\makeatletter
\providecommand \@ifxundefined [1]{%
 \@ifx{#1\undefined}
}%
\providecommand \@ifnum [1]{%
 \ifnum #1\expandafter \@firstoftwo
 \else \expandafter \@secondoftwo
 \fi
}%
\providecommand \@ifx [1]{%
 \ifx #1\expandafter \@firstoftwo
 \else \expandafter \@secondoftwo
 \fi
}%
\providecommand \natexlab [1]{#1}%
\providecommand \enquote  [1]{``#1''}%
\providecommand \bibnamefont  [1]{#1}%
\providecommand \bibfnamefont [1]{#1}%
\providecommand \citenamefont [1]{#1}%
\providecommand \href@noop [0]{\@secondoftwo}%
\providecommand \href [0]{\begingroup \@sanitize@url \@href}%
\providecommand \@href[1]{\@@startlink{#1}\@@href}%
\providecommand \@@href[1]{\endgroup#1\@@endlink}%
\providecommand \@sanitize@url [0]{\catcode `\\12\catcode `\$12\catcode
  `\&12\catcode `\#12\catcode `\^12\catcode `\_12\catcode `\%12\relax}%
\providecommand \@@startlink[1]{}%
\providecommand \@@endlink[0]{}%
\providecommand \url  [0]{\begingroup\@sanitize@url \@url }%
\providecommand \@url [1]{\endgroup\@href {#1}{\urlprefix }}%
\providecommand \urlprefix  [0]{URL }%
\providecommand \Eprint [0]{\href }%
\providecommand \doibase [0]{http://dx.doi.org/}%
\providecommand \selectlanguage [0]{\@gobble}%
\providecommand \bibinfo  [0]{\@secondoftwo}%
\providecommand \bibfield  [0]{\@secondoftwo}%
\providecommand \translation [1]{[#1]}%
\providecommand \BibitemOpen [0]{}%
\providecommand \bibitemStop [0]{}%
\providecommand \bibitemNoStop [0]{.\EOS\space}%
\providecommand \EOS [0]{\spacefactor3000\relax}%
\providecommand \BibitemShut  [1]{\csname bibitem#1\endcsname}%
\let\auto@bib@innerbib\@empty
%</preamble>
\bibitem [{\citenamefont {Awschalom}\ and\ \citenamefont
  {Samarth}(1999)}]{Awschalom}%
  \BibitemOpen
  \bibfield  {author} {\bibinfo {author} {\bibfnamefont {D.}~\bibnamefont
  {Awschalom}}\ and\ \bibinfo {author} {\bibfnamefont {N.}~\bibnamefont
  {Samarth}},\ }\href {\doibase
  http://dx.doi.org/10.1016/S0304-8853(99)00424-2} {\bibfield  {journal}
  {\bibinfo  {journal} {Journal of Magnetism and Magnetic Materials}\ }\textbf
  {\bibinfo {volume} {200}},\ \bibinfo {pages} {130 } (\bibinfo {year}
  {1999})}\BibitemShut {NoStop}%
\bibitem [{\citenamefont {Awschalom}\ and\ \citenamefont
  {Flatt\'{e}}(2007)}]{Awschalom07}%
  \BibitemOpen
  \bibfield  {author} {\bibinfo {author} {\bibfnamefont {D.}~\bibnamefont
  {Awschalom}}\ and\ \bibinfo {author} {\bibfnamefont {M.}~\bibnamefont
  {Flatt\'{e}}},\ }\href@noop {} {\bibfield  {journal} {\bibinfo  {journal}
  {Nature Physics}\ }\textbf {\bibinfo {volume} {3}},\ \bibinfo {pages} {153 }
  (\bibinfo {year} {2007})}\BibitemShut {NoStop}%
\bibitem [{\citenamefont {Wolf}\ \emph {et~al.}(2001)\citenamefont {Wolf},
  \citenamefont {Awschalom}, \citenamefont {Buhrman}, \citenamefont {Daughton},
  \citenamefont {von Molnár}, \citenamefont {Roukes}, \citenamefont
  {Chtchelkanova},\ and\ \citenamefont {Treger}}]{Spintronics}%
  \BibitemOpen
  \bibfield  {author} {\bibinfo {author} {\bibfnamefont {S.~A.}\ \bibnamefont
  {Wolf}}, \bibinfo {author} {\bibfnamefont {D.~D.}\ \bibnamefont {Awschalom}},
  \bibinfo {author} {\bibfnamefont {R.~A.}\ \bibnamefont {Buhrman}}, \bibinfo
  {author} {\bibfnamefont {J.~M.}\ \bibnamefont {Daughton}}, \bibinfo {author}
  {\bibfnamefont {S.}~\bibnamefont {von Molnár}}, \bibinfo {author}
  {\bibfnamefont {M.~L.}\ \bibnamefont {Roukes}}, \bibinfo {author}
  {\bibfnamefont {A.~Y.}\ \bibnamefont {Chtchelkanova}}, \ and\ \bibinfo
  {author} {\bibfnamefont {D.~M.}\ \bibnamefont {Treger}},\ }\href {\doibase
  10.1126/science.1065389} {\bibfield  {journal} {\bibinfo  {journal}
  {Science}\ }\textbf {\bibinfo {volume} {294}},\ \bibinfo {pages} {1488}
  (\bibinfo {year} {2001})}\BibitemShut {NoStop}%
\bibitem [{\citenamefont {\ifmmode \check{Z}\else
  \v{Z}\fi{}uti\ifmmode~\acute{c}\else \'{c}\fi{}}\ \emph
  {et~al.}(2004)\citenamefont {\ifmmode \check{Z}\else
  \v{Z}\fi{}uti\ifmmode~\acute{c}\else \'{c}\fi{}}, \citenamefont {Fabian},\
  and\ \citenamefont {Das~Sarma}}]{Zutic}%
  \BibitemOpen
  \bibfield  {author} {\bibinfo {author} {\bibfnamefont {I.}~\bibnamefont
  {\ifmmode \check{Z}\else \v{Z}\fi{}uti\ifmmode~\acute{c}\else \'{c}\fi{}}},
  \bibinfo {author} {\bibfnamefont {J.}~\bibnamefont {Fabian}}, \ and\ \bibinfo
  {author} {\bibfnamefont {S.}~\bibnamefont {Das~Sarma}},\ }\href {\doibase
  10.1103/RevModPhys.76.323} {\bibfield  {journal} {\bibinfo  {journal} {Rev.
  Mod. Phys.}\ }\textbf {\bibinfo {volume} {76}},\ \bibinfo {pages} {323}
  (\bibinfo {year} {2004})}\BibitemShut {NoStop}%
\bibitem [{\citenamefont {MacDonald}\ \emph {et~al.}(2005)\citenamefont
  {MacDonald}, \citenamefont {Schiffer},\ and\ \citenamefont
  {Samarth}}]{MacDonald}%
  \BibitemOpen
  \bibfield  {author} {\bibinfo {author} {\bibfnamefont {A.}~\bibnamefont
  {MacDonald}}, \bibinfo {author} {\bibfnamefont {P.}~\bibnamefont {Schiffer}},
  \ and\ \bibinfo {author} {\bibfnamefont {N.}~\bibnamefont {Samarth}},\
  }\href@noop {} {\bibfield  {journal} {\bibinfo  {journal} {Nature Materials}\
  }\textbf {\bibinfo {volume} {4}},\ \bibinfo {pages} {195 } (\bibinfo {year}
  {2005})}\BibitemShut {NoStop}%
\bibitem [{\citenamefont {Dietl}\ \emph {et~al.}(2000)\citenamefont {Dietl},
  \citenamefont {Ohno}, \citenamefont {Matsukura}, \citenamefont {Cibert},\
  and\ \citenamefont {Ferrand}}]{DOM}%
  \BibitemOpen
  \bibfield  {author} {\bibinfo {author} {\bibfnamefont {T.}~\bibnamefont
  {Dietl}}, \bibinfo {author} {\bibfnamefont {H.}~\bibnamefont {Ohno}},
  \bibinfo {author} {\bibfnamefont {F.}~\bibnamefont {Matsukura}}, \bibinfo
  {author} {\bibfnamefont {J.}~\bibnamefont {Cibert}}, \ and\ \bibinfo {author}
  {\bibfnamefont {D.}~\bibnamefont {Ferrand}},\ }\href {\doibase
  10.1126/science.287.5455.1019} {\bibfield  {journal} {\bibinfo  {journal}
  {Science}\ }\textbf {\bibinfo {volume} {287}},\ \bibinfo {pages} {1019}
  (\bibinfo {year} {2000})}\BibitemShut {NoStop}%
\bibitem [{\citenamefont {Wolff}\ and\ \citenamefont {Warnock}(1984)}]{Wolff}%
  \BibitemOpen
  \bibfield  {author} {\bibinfo {author} {\bibfnamefont {P.~A.}\ \bibnamefont
  {Wolff}}\ and\ \bibinfo {author} {\bibfnamefont {J.}~\bibnamefont
  {Warnock}},\ }\href {\doibase http://dx.doi.org/10.1063/1.333642} {\bibfield
  {journal} {\bibinfo  {journal} {Journal of Applied Physics}\ }\textbf
  {\bibinfo {volume} {55}},\ \bibinfo {pages} {2300} (\bibinfo {year}
  {1984})}\BibitemShut {NoStop}%
\bibitem [{\citenamefont {Morandi}\ \emph {et~al.}(2009)\citenamefont
  {Morandi}, \citenamefont {Hervieux},\ and\ \citenamefont
  {Manfredi}}]{Morandi2009}%
  \BibitemOpen
  \bibfield  {author} {\bibinfo {author} {\bibfnamefont {O.}~\bibnamefont
  {Morandi}}, \bibinfo {author} {\bibfnamefont {P.-A.}\ \bibnamefont
  {Hervieux}}, \ and\ \bibinfo {author} {\bibfnamefont {G.}~\bibnamefont
  {Manfredi}},\ }\href {http://stacks.iop.org/1367-2630/11/i=7/a=073010}
  {\bibfield  {journal} {\bibinfo  {journal} {New Journal of Physics}\ }\textbf
  {\bibinfo {volume} {11}},\ \bibinfo {pages} {073010} (\bibinfo {year}
  {2009})}\BibitemShut {NoStop}%
\bibitem [{\citenamefont {Jiang}\ \emph {et~al.}(2009)\citenamefont {Jiang},
  \citenamefont {Zhou}, \citenamefont {Korn}, \citenamefont {Sch\"uller},\ and\
  \citenamefont {Wu}}]{Wu09}%
  \BibitemOpen
  \bibfield  {author} {\bibinfo {author} {\bibfnamefont {J.~H.}\ \bibnamefont
  {Jiang}}, \bibinfo {author} {\bibfnamefont {Y.}~\bibnamefont {Zhou}},
  \bibinfo {author} {\bibfnamefont {T.}~\bibnamefont {Korn}}, \bibinfo {author}
  {\bibfnamefont {C.}~\bibnamefont {Sch\"uller}}, \ and\ \bibinfo {author}
  {\bibfnamefont {M.~W.}\ \bibnamefont {Wu}},\ }\href {\doibase
  10.1103/PhysRevB.79.155201} {\bibfield  {journal} {\bibinfo  {journal} {Phys.
  Rev. B}\ }\textbf {\bibinfo {volume} {79}},\ \bibinfo {pages} {155201}
  (\bibinfo {year} {2009})}\BibitemShut {NoStop}%
\bibitem [{\citenamefont {K\"onig}\ \emph {et~al.}(2000)\citenamefont
  {K\"onig}, \citenamefont {Merkulov}, \citenamefont {Yakovlev}, \citenamefont
  {Ossau}, \citenamefont {Ryabchenko}, \citenamefont {Kutrowski}, \citenamefont
  {Wojtowicz}, \citenamefont {Karczewski},\ and\ \citenamefont
  {Kossut}}]{Koenig}%
  \BibitemOpen
  \bibfield  {author} {\bibinfo {author} {\bibfnamefont {B.}~\bibnamefont
  {K\"onig}}, \bibinfo {author} {\bibfnamefont {I.~A.}\ \bibnamefont
  {Merkulov}}, \bibinfo {author} {\bibfnamefont {D.~R.}\ \bibnamefont
  {Yakovlev}}, \bibinfo {author} {\bibfnamefont {W.}~\bibnamefont {Ossau}},
  \bibinfo {author} {\bibfnamefont {S.~M.}\ \bibnamefont {Ryabchenko}},
  \bibinfo {author} {\bibfnamefont {M.}~\bibnamefont {Kutrowski}}, \bibinfo
  {author} {\bibfnamefont {T.}~\bibnamefont {Wojtowicz}}, \bibinfo {author}
  {\bibfnamefont {G.}~\bibnamefont {Karczewski}}, \ and\ \bibinfo {author}
  {\bibfnamefont {J.}~\bibnamefont {Kossut}},\ }\href {\doibase
  10.1103/PhysRevB.61.16870} {\bibfield  {journal} {\bibinfo  {journal} {Phys.
  Rev. B}\ }\textbf {\bibinfo {volume} {61}},\ \bibinfo {pages} {16870}
  (\bibinfo {year} {2000})}\BibitemShut {NoStop}%
\bibitem [{\citenamefont {Cywi\ifmmode~\acute{n}\else \'{n}\fi{}ski}\ and\
  \citenamefont {Sham}(2007)}]{Cywinski}%
  \BibitemOpen
  \bibfield  {author} {\bibinfo {author} {\bibfnamefont {L.}~\bibnamefont
  {Cywi\ifmmode~\acute{n}\else \'{n}\fi{}ski}}\ and\ \bibinfo {author}
  {\bibfnamefont {L.~J.}\ \bibnamefont {Sham}},\ }\href {\doibase
  10.1103/PhysRevB.76.045205} {\bibfield  {journal} {\bibinfo  {journal} {Phys.
  Rev. B}\ }\textbf {\bibinfo {volume} {76}},\ \bibinfo {pages} {045205}
  (\bibinfo {year} {2007})}\BibitemShut {NoStop}%
\bibitem [{\citenamefont {Semenov}(2003)}]{Semenov}%
  \BibitemOpen
  \bibfield  {author} {\bibinfo {author} {\bibfnamefont {Y.~G.}\ \bibnamefont
  {Semenov}},\ }\href {\doibase 10.1103/PhysRevB.67.115319} {\bibfield
  {journal} {\bibinfo  {journal} {Phys. Rev. B}\ }\textbf {\bibinfo {volume}
  {67}},\ \bibinfo {pages} {115319} (\bibinfo {year} {2003})}\BibitemShut
  {NoStop}%
\bibitem [{\citenamefont {Kapetanakis}\ \emph {et~al.}(2012)\citenamefont
  {Kapetanakis}, \citenamefont {Wang},\ and\ \citenamefont
  {Perakis}}]{Perakis2012}%
  \BibitemOpen
  \bibfield  {author} {\bibinfo {author} {\bibfnamefont {M.~D.}\ \bibnamefont
  {Kapetanakis}}, \bibinfo {author} {\bibfnamefont {J.}~\bibnamefont {Wang}}, \
  and\ \bibinfo {author} {\bibfnamefont {I.~E.}\ \bibnamefont {Perakis}},\
  }\href {\doibase 10.1364/JOSAB.29.000A95} {\bibfield  {journal} {\bibinfo
  {journal} {J. Opt. Soc. Am. B}\ }\textbf {\bibinfo {volume} {29}},\ \bibinfo
  {pages} {A95} (\bibinfo {year} {2012})}\BibitemShut {NoStop}%
\bibitem [{\citenamefont {Morandi}(2011)}]{Morandi2011}%
  \BibitemOpen
  \bibfield  {author} {\bibinfo {author} {\bibfnamefont {O.}~\bibnamefont
  {Morandi}},\ }\href {\doibase 10.1103/PhysRevB.83.224428} {\bibfield
  {journal} {\bibinfo  {journal} {Phys. Rev. B}\ }\textbf {\bibinfo {volume}
  {83}},\ \bibinfo {pages} {224428} (\bibinfo {year} {2011})}\BibitemShut
  {NoStop}%
\bibitem [{\citenamefont {Thurn}\ and\ \citenamefont {Axt}(2012)}]{Thurn:12}%
  \BibitemOpen
  \bibfield  {author} {\bibinfo {author} {\bibfnamefont {C.}~\bibnamefont
  {Thurn}}\ and\ \bibinfo {author} {\bibfnamefont {V.~M.}\ \bibnamefont
  {Axt}},\ }\href {\doibase 10.1103/PhysRevB.85.165203} {\bibfield  {journal}
  {\bibinfo  {journal} {Phys. Rev. B}\ }\textbf {\bibinfo {volume} {85}},\
  \bibinfo {pages} {165203} (\bibinfo {year} {2012})}\BibitemShut {NoStop}%
\bibitem [{\citenamefont {Thurn}\ \emph
  {et~al.}(2013{\natexlab{a}})\citenamefont {Thurn}, \citenamefont {Cygorek},
  \citenamefont {Axt},\ and\ \citenamefont {Kuhn}}]{Thurn:13_1}%
  \BibitemOpen
  \bibfield  {author} {\bibinfo {author} {\bibfnamefont {C.}~\bibnamefont
  {Thurn}}, \bibinfo {author} {\bibfnamefont {M.}~\bibnamefont {Cygorek}},
  \bibinfo {author} {\bibfnamefont {V.~M.}\ \bibnamefont {Axt}}, \ and\
  \bibinfo {author} {\bibfnamefont {T.}~\bibnamefont {Kuhn}},\ }\href {\doibase
  10.1103/PhysRevB.87.205301} {\bibfield  {journal} {\bibinfo  {journal} {Phys.
  Rev. B}\ }\textbf {\bibinfo {volume} {87}},\ \bibinfo {pages} {205301}
  (\bibinfo {year} {2013}{\natexlab{a}})}\BibitemShut {NoStop}%
\bibitem [{\citenamefont {Thurn}\ \emph
  {et~al.}(2013{\natexlab{b}})\citenamefont {Thurn}, \citenamefont {Cygorek},
  \citenamefont {Axt},\ and\ \citenamefont {Kuhn}}]{Thurn:13_2}%
  \BibitemOpen
  \bibfield  {author} {\bibinfo {author} {\bibfnamefont {C.}~\bibnamefont
  {Thurn}}, \bibinfo {author} {\bibfnamefont {M.}~\bibnamefont {Cygorek}},
  \bibinfo {author} {\bibfnamefont {V.~M.}\ \bibnamefont {Axt}}, \ and\
  \bibinfo {author} {\bibfnamefont {T.}~\bibnamefont {Kuhn}},\ }\href {\doibase
  10.1103/PhysRevB.88.161302} {\bibfield  {journal} {\bibinfo  {journal} {Phys.
  Rev. B}\ }\textbf {\bibinfo {volume} {88}},\ \bibinfo {pages} {161302}
  (\bibinfo {year} {2013}{\natexlab{b}})}\BibitemShut {NoStop}%
\bibitem [{\citenamefont {Zener}(1951)}]{Zener}%
  \BibitemOpen
  \bibfield  {author} {\bibinfo {author} {\bibfnamefont {C.}~\bibnamefont
  {Zener}},\ }\href {\doibase 10.1103/PhysRev.81.440} {\bibfield  {journal}
  {\bibinfo  {journal} {Phys. Rev.}\ }\textbf {\bibinfo {volume} {81}},\
  \bibinfo {pages} {440} (\bibinfo {year} {1951})}\BibitemShut {NoStop}%
\bibitem [{\citenamefont {Kossut}(1988)}]{semicond_semimet}%
  \BibitemOpen
  \bibfield  {author} {\bibinfo {author} {\bibfnamefont {J.}~\bibnamefont
  {Kossut}},\ }\href@noop {} {\emph {\bibinfo {title} {Diluted Magnetic
  Semiconductors}}},\ edited by\ \bibinfo {editor} {\bibfnamefont
  {J.}~\bibnamefont {Furdyna}}\ and\ \bibinfo {editor} {\bibfnamefont
  {J.}~\bibnamefont {Kossut}},\ \bibinfo {series} {Semiconductors and
  Semimetals}, Vol.~\bibinfo {volume} {25}\ (\bibinfo  {publisher} {Academic
  Press},\ \bibinfo {address} {San Diego},\ \bibinfo {year} {1988})\ p.\
  \bibinfo {pages} {185}\BibitemShut {NoStop}%
\bibitem [{Note1()}]{Note1}%
  \BibitemOpen
  \bibinfo {note} {For lower dimensional systems this crude approximation leads
  to a divergence of the frequency renormalization at $k\rightarrow k_1$. This
  fact supports the findings of Refs.~\protect \rev@citealpnum
  {Thurn:13_1,Thurn:13_2} that the Markov limit is not a good approximation in
  systems with dimensions lower than 3.}\BibitemShut {Stop}%
\bibitem [{\citenamefont {H\"ubner}\ and\ \citenamefont
  {Oestreich}(2008)}]{SpinPhysSemi5}%
  \BibitemOpen
  \bibfield  {author} {\bibinfo {author} {\bibfnamefont {J.}~\bibnamefont
  {H\"ubner}}\ and\ \bibinfo {author} {\bibfnamefont {M.}~\bibnamefont
  {Oestreich}},\ }\href@noop {} {\emph {\bibinfo {title} {Spin Physics in
  Semiconductors}}},\ edited by\ \bibinfo {editor} {\bibfnamefont {M.~I.}\
  \bibnamefont {Dyakonov}},\ \bibinfo {series} {Springer Series in Solid-State
  Sciences}, Vol.\ \bibinfo {volume} {157}\ (\bibinfo  {publisher} {Springer},\
  \bibinfo {year} {2008})\ pp.\ \bibinfo {pages} {115--134}\BibitemShut
  {NoStop}%
\bibitem [{\citenamefont {Bychkov}\ and\ \citenamefont
  {Rashba}(1984)}]{Rashba}%
  \BibitemOpen
  \bibfield  {author} {\bibinfo {author} {\bibfnamefont {Y.~A.}\ \bibnamefont
  {Bychkov}}\ and\ \bibinfo {author} {\bibfnamefont {E.~I.}\ \bibnamefont
  {Rashba}},\ }\href {http://stacks.iop.org/0022-3719/17/i=33/a=015} {\bibfield
   {journal} {\bibinfo  {journal} {Journal of Physics C: Solid State Physics}\
  }\textbf {\bibinfo {volume} {17}},\ \bibinfo {pages} {6039} (\bibinfo {year}
  {1984})}\BibitemShut {NoStop}%
\bibitem [{\citenamefont {Bir}\ \emph {et~al.}(1975)\citenamefont {Bir},
  \citenamefont {Aronov},\ and\ \citenamefont {Pikus}}]{BAP}%
  \BibitemOpen
  \bibfield  {author} {\bibinfo {author} {\bibfnamefont {G.~L.}\ \bibnamefont
  {Bir}}, \bibinfo {author} {\bibfnamefont {A.}~\bibnamefont {Aronov}}, \ and\
  \bibinfo {author} {\bibfnamefont {G.~E.}\ \bibnamefont {Pikus}},\ }\href@noop
  {} {\bibfield  {journal} {\bibinfo  {journal} {JETP}\ }\textbf {\bibinfo
  {volume} {42}},\ \bibinfo {pages} {705} (\bibinfo {year} {1975})}\BibitemShut
  {NoStop}%
\bibitem [{\citenamefont {Wu}\ \emph {et~al.}(2010)\citenamefont {Wu},
  \citenamefont {Jiang},\ and\ \citenamefont {Weng}}]{WuReview}%
  \BibitemOpen
  \bibfield  {author} {\bibinfo {author} {\bibfnamefont {M.}~\bibnamefont
  {Wu}}, \bibinfo {author} {\bibfnamefont {J.}~\bibnamefont {Jiang}}, \ and\
  \bibinfo {author} {\bibfnamefont {M.}~\bibnamefont {Weng}},\ }\href {\doibase
  http://dx.doi.org/10.1016/j.physrep.2010.04.002} {\bibfield  {journal}
  {\bibinfo  {journal} {Physics Reports}\ }\textbf {\bibinfo {volume} {493}},\
  \bibinfo {pages} {61 } (\bibinfo {year} {2010})}\BibitemShut {NoStop}%
\end{thebibliography}

\end{document}